\documentclass[aps,pra,10pt,twocolumn,superscriptaddress,showpacs,floatfix]{revtex4-2}
\usepackage{amsmath,amssymb,bm,graphicx,booktabs}
\usepackage[T1]{fontenc}
\usepackage[nopatch=footnote]{microtype}
\usepackage{url,capt-of}
\AtBeginEnvironment{thebibliography}{\raggedright}
\graphicspath{{figs/}}
\newcommand{\av}[1]{\langle #1\rangle}
\newcommand{\F}{F_{\max}}
\newcommand{\fd}{F_{\mathrm{dip}}}
\newcommand{\Pe}{P_{\mathrm{ex}}}
\begin{document}
\title{Finite-size spin fluctuations and fixed-index sweeps in a ferromagnetic spinor condensate}
\author{Yi Ding}
\affiliation{Babes Bolyai University}
\author{Yixiao Huang}
\affiliation{Zhejiang University of Science and Technology}
\author{Junyan Luo}
\affiliation{Zhejiang University of Science and Technology}
\begin{abstract}
We find a dip in the optimized rotation quantum Fisher information of a zero-magnetization ferromagnetic spin-1 Bose-Einstein condensate near the excited-state quantum phase transition. Its displacement from $\epsilon_c=-q$ decreases on an inverse-size scale. At $q=0.6,0.8,1.0,1.2$, the dip value of $F_{\max}$ equals $2\langle S^2\rangle$ for $N$ up to $3840$, linking it to total-spin fluctuations. A semiclassical orbit average normalized by the classical density of states describes the dip over $N=120$--$3840$ with maximum relative residual below $0.028$; a small power law fits comparably. Inserting the measured energy offsets directly into the semiclassical formula leaves residuals of $17$--$30\%$, so $\kappa$ acts as an effective cutoff. Among retained states, spin-nematic squeezing occurs at the spectral edges. Separately, fixed-index excited-state sweeps have half-excitation times between $11.2$ and $13.1$ times their ground-state counterparts, with local size exponents increasing across the four sizes.
\end{abstract}
\maketitle

\section{Introduction}
Excited-state quantum phase transitions (ESQPTs) extend critical behavior into the spectrum of a many-body Hamiltonian. Classical stationary points and separatrices provide a framework for understanding singular level densities and eigenstate observables in collective systems \cite{Caprio2008,Cejnar2021}. Early semiclassical work linked the spectral singularity of the Lipkin--Meshkov--Glick (LMG) model to its separatrix and diverging period \cite{Leyvraz2005}. Subsequent studies connected excited-level flow to classical stationary points \cite{Cejnar2008} and established how the singularity changes with the number of collective degrees of freedom \cite{Stransky2014}. Critical eigenstates also exhibit basis-dependent localization in two-level pairing models \cite{Santos2015}.

Spin-dependent collisions support polar and ferromagnetic phases in spin-1 Bose-Einstein condensates (BECs) \cite{Ho1998,Ohmi1998}. Coherent spin-mixing oscillations were observed in ferromagnetic rubidium \cite{Chang2005} and antiferromagnetic sodium \cite{Black2007}. Quenches across the ferromagnetic ground-state transition produced spatial spin domains \cite{Sadler2006}, while tightly confined condensates exhibited fluctuations near a hyperbolic fixed point and non-Gaussian spin-population distributions \cite{Gerving2012}. Ferromagnetic spin-1 BECs also realize an excited-state phase diagram: Feldmann \emph{et al.} distinguished phases by a trajectory winding number and proposed an interferometric order parameter \cite{Feldmann2021}. This phase diagram has since been mapped experimentally using an interferometric order parameter \cite{MeyerHoppe2023}.

Spin squeezing connects reduced collective-spin fluctuations to improved spectroscopic precision \cite{Kitagawa1993,Wineland1994}. In spin-1 ensembles, spin and nematic operators form an SU(3) algebra with two inequivalent classes of SU(2) squeezing \cite{Yukawa2013}. Spin-nematic quadrature squeezing has been observed experimentally in a spin-1 condensate \cite{Hamley2012}. Interferometry beyond the classical precision limit has used twin matter waves in the F=2 manifold \cite{Lucke2011} and spin-1 Dicke states \cite{Zou2018}. Three-mode spin-mixing interferometry also provides a route to sub-shot-noise phase estimation \cite{Gabbrielli2015}.

In antiferromagnetic spinor gases, dynamical transitions have been connected experimentally to the highest-energy phase boundary within a fixed-magnetization sector \cite{Tian2020}. In quench dynamics, squeezing has been used to identify transitions associated with the ground state and the highest energy level of an antiferromagnetic spinor condensate \cite{Huang2022}. Related work connects spinor dynamics to critical sensing \cite{Zhou2023} and to semiclassical phase-space structure \cite{Niu2023}. Semiclassical periodic-orbit averages have been related to eigenstate expectation values in the LMG and Tavis--Cummings models \cite{Engelhardt2015}. In the spinor condensate, eigenstate localization near the ESQPT has also been characterized through dips in Fock-basis fractal dimensions \cite{Niu2024}.

Ground-state critical crossings connect excitation production to the ramp rate through quantum Kibble--Zurek arguments \cite{Zurek2005,Polkovnikov2005}. The exact Ising solution distinguishes defect-density scaling from the ground-state survival probability of a finite chain \cite{Dziarmaga2005}. In ferromagnetic spinor gases, mean-field theory describes delayed magnetization growth during a slow crossing of the polar transition \cite{Damski2007}, and experiments have tested the quantum Kibble--Zurek mechanism \cite{Anquez2016}. Adiabatic spinor protocols were proposed to prepare entanglement near Dicke states \cite{Zhang2013} and subsequently realized through ground-state transition crossings \cite{Luo2017}. Measurements of the finite-size energy gap have also guided adiabatic passage across the ground-state transition \cite{Hoang2016}.

Long-lived squeezed ground states have been realized experimentally \cite{Xin2023}, and nonadiabatic protocols have been proposed for preparing time-stationary squeezed states \cite{Xin2022}. These results motivate studying static eigenstate fluctuations and fixed-index slow sweeps.

We ask two questions: what signal does the ESQPT leave in eigenstate fluctuations, and how long does a slow sweep take to follow an excited level through it? We find (i) a dip near $\epsilon_c=-q$ whose $\F$ equals $2\av{S^2}$ at $q=0.6,0.8,1.0,1.2$ for $N$ up to $3840$, alongside spin-nematic squeezing at the spectral edges; (ii) comparable semiclassical and small-power descriptions of the dip over $N=120$--$3840$; and (iii) excited-state half-excitation times between $11.2$ and $13.1$ times the ground-state values. These findings connect the static spectral feature to its classical orbit average and quantify the separate dynamical problem of following an eigenstate.

\section{Model and ESQPT}
The single-mode approximation uses one spatial wavefunction for all three spin components \cite{Yi2002,Kawaguchi2012}. It reduces the spin-dependent interaction to a collective-spin Hamiltonian \cite{Law1998}. The zero-magnetization ferromagnetic condensate is described by the single-mode Hamiltonian \cite{Huang2022,Feldmann2021}
\begin{equation}
 H(q)=\frac{c}{2N}S^2-q n_0,\qquad c=-1,\quad \hbar=1 ,
 \label{eq:model}
\end{equation}
in the magnetization sector $M=n_1-n_{-1}=0$. This sector is natural for experiments initialized in the polar state with all atoms in $m=0$, because that state has $M=0$ and the Hamiltonian conserves magnetization \cite{Feldmann2021}. Energy and time are measured in $|c|$ and $\hbar/|c|$, respectively. For even $N$, the basis
$|k\rangle=|k,N-2k,k\rangle$, $k=0,\ldots,N/2$, gives
\begin{align}
 H_{kk}&=\frac{c}{2N}\big[2n_0(2k+1)+2k\big]-q n_0,\\
 H_{k,k+1}&=\frac{c}{N}(k+1)\sqrt{(N-2k)(N-2k-1)},
 \label{eq:matrix}
\end{align}
where $n_0=N-2k$. The levels are ordered increasingly as $E_m$.
Mean-field spin mixing follows a nonrigid pendulum with a separatrix \cite{Zhang2005}. Writing $\epsilon=E/N$, $\rho=n_0/N$, and
$\phi=\phi_0-(\phi_1+\phi_{-1})/2$, the classical energy is
\begin{equation}
 h(\rho,\phi)=-q\rho-2\rho(1-\rho)\cos^2\phi .
 \label{eq:classical}
\end{equation}
The polar stationary point has $\epsilon_c=-q$, giving $E_c=-qN$ for $0<q<2$; the positive-field ground-state critical point is $q=2$ \cite{Feldmann2021}.

To compare conventions explicitly, the Hamiltonian density $\hat h_{\rm F}$ of Ref.~\cite{Feldmann2021}, Eq.~(1), obeys
\begin{equation}
 \hat h_{\rm F}=H/N+q/2-c/(2N).
 \label{eq:mapping}
\end{equation}
Thus its field variable is $\xi_{\rm F}=q/(2|c|)$, and its energy variable at $c=-1$ is $\eta_N=\epsilon+q/2+1/(2N)$, with classical limit $\eta=\epsilon+q/2$. Its critical line $\eta_*=-|\xi_{\rm F}|$ maps to $\epsilon_c=-q$ for positive $q$. Its broken-axisymmetry excited-state region, BA$'$, lies below this line and has winding number zero; the squeezing observable is defined separately below.

\par\medskip\noindent\begin{minipage}{\columnwidth}

\includegraphics[width=\columnwidth]{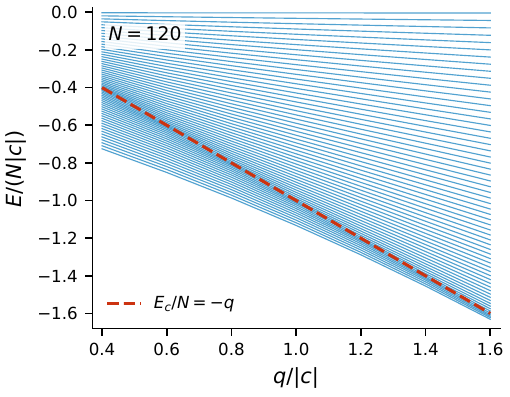}
\captionof{figure}{Zero-magnetization spectrum at $N=120$. Lines connect all $61$ levels at the seven fields $q=0.4,0.6,\ldots,1.6$; segments are guides to the eye. The dashed line is $E_c/N=-q$.}
\label{fig:spectrum}

\end{minipage}
\par\medskip

In the variable $\eta_N$ the offset from the critical line is $\delta+1/(2N)$, where $\delta=\epsilon+q$ is the offset used below.

\section{Eigenstate fluctuations at the critical energy}
The spin and nematic operators follow the normalization of Refs.~\cite{Huang2015,Huang2022}; the operators are given in Appendix~\ref{app:operators}. In an $M=0$ eigenstate, $\av{S_x}=\av{Q_{yz}}=0$, and the covariance combinations are
\begin{align}
 A&=\av{S_x^2+Q_{yz}^2},\quad B=\av{S_x^2-Q_{yz}^2},\\
 C&=\av{S_xQ_{yz}+Q_{yz}S_x},\quad R=\sqrt{B^2+C^2}.
\end{align}
For a fixed pure state, the quantum Fisher information (QFI) for estimating a rotation angle is four times the generator variance \cite{Braunstein1994}. Optimizing over normalized linear combinations of generators is a covariance-matrix eigenvalue problem \cite{Hyllus2012}. Using the spin-nematic squeezing normalization \cite{Huang2015,Huang2022}, the minimum variance and the QFI optimized over the displayed spin-nematic plane are
\begin{align}
 \xi_x^2&=\frac{A-R}{|\av{Q_+}|},\qquad Q_+=Q_{zz}-Q_{yy},\\
 \F&=4\max_\theta\av{[\Delta(\cos\theta S_x+\sin\theta Q_{yz})]^2}
      =2(A+R).
 \label{eq:qfi}
\end{align}
The polar state gives the reference values $\xi_x^2=1$ and $\F/N=4$. Squeezing summaries retain states with $|\av{Q_+}|\geq0.1N$, while QFI profiles include every eigenstate.

\subsection{The Fisher-information and total-spin dips}
The critical-energy feature is a dip, while the global $\F$ maximum lies at the ground-state end for $N=60,120,240$ and $q=0.4,0.6,0.8,1.0,1.2$ (Fig.~\ref{fig:profiles}). At $q=1.4,1.6$ the global maximum moves into the spectrum. At $q=1$, the ground-state values of $\F/N^2$ are $1.522$, $1.511$, and $1.506$, respectively; the BA$'$ profiles exhibit a finite-size $N^2$ scale.

The dip approaches $\epsilon_c=-q$: its median absolute displacement decreases from $0.0121$ to $0.0070$ to $0.0031$ for $N=60,120,240$. We define the dip state by minimizing $\F$ over $|\epsilon+q|<0.06$ and set $\fd=\F(\epsilon_{\rm dip})/N^2$. At $N=240$, the offsets at $q=0.4,0.6,0.8,1.0,1.2$ are $-0.00179$, $-0.00159$, $-0.00392$, $-0.00312$, and $-0.00405$. In the extended four-field dataset of Sec.~\ref{sec:semiclassical}, $N|\epsilon_{\rm dip}+q|$ ranges from $0.3283$ to $1.440$, consistent with an inverse-size displacement scale modulated by level discreteness.

At the dip we find
\begin{equation}
 \F=2\av{S^2}
 \label{eq:identity}
\end{equation}
for sizes $N=60$, $120$, $240$, $480$, $960$, $1920$, and $3840$ at $q=0.6,0.8,1.0,1.2$. At $N=240,q=0.4$, the two quantities separate: $\fd=0.4125$ and $2\av{S^2}/N^2=0.3080$. At $q=0.6$, states toward the upper part of the energy window have maximum relative deviations $|\F/(2\av{S^2})-1|$ of $0.3402$, $0.3443$, and $0.3438$ for $N=960,1920,3840$, respectively. Appendix~\ref{app:operators} gives the operator construction and numerical accuracy at the minima.

The two dips have equal relative depth at $q=1.0,1.2$, while the QFI dip is deeper by $0.067$ and $0.082$ at $q=0.4$ and $0.6$ (Table~\ref{tab:depth}). Here $D_F=1-\fd/\bar F$, with $\bar F$ averaging $\F/N^2$ over the two bands $0.06\leq|\epsilon+q|\leq0.12$, each containing at least three levels. The total-spin depth $D_S$ uses the same construction for $\av{S^2}/N^2$, with its own window minimum. At $q=0.8$ the depth difference is $0.0119$.

\par\medskip\noindent\begin{minipage}{\columnwidth}

\captionof{table}{Dip depths at $N=240$. For the three-size comparison, the $N=60$ bands contain $(2,2)$, $(3,2)$, and $(2,2)$ states at $q=0.8,1,1.2$, respectively, leaving these points outside the three-states-per-band sample.}\label{tab:depth}
\begin{ruledtabular}\begin{tabular}{rrrr}
$q$&$D_F$&$D_S$&$D_F-D_S$\\
0.4&0.606964&0.540309&0.066656\\
0.6&0.603882&0.521705&0.082177\\
0.8&0.533617&0.521751&0.011866\\
1.0&0.527332&0.527332&0\\
1.2&0.491583&0.491583&0
\end{tabular}\end{ruledtabular}

\end{minipage}
\par\medskip

\onecolumngrid
\noindent\begin{minipage}{\textwidth}

\includegraphics[width=\textwidth]{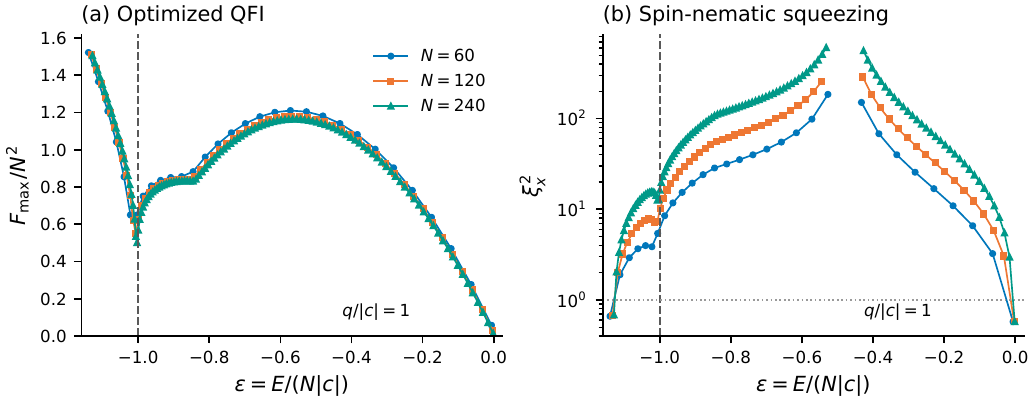}
\captionof{figure}{Eigenstate profiles at $q=1$ for $N=60,120,240$: (a) $\F/N^2$ and (b) $\xi_x^2$. Vertical dashed lines mark $\epsilon_c=-1$; the horizontal dotted line in (b) marks $\xi_x^2=1$. States with $|\av{Q_+}|<0.1N$ are omitted from (b), leaving gaps. Curves connect eigenstate values.}
\label{fig:profiles}

\end{minipage}
\par\medskip
\twocolumngrid

\subsection{Spin-nematic squeezing at the spectral edges}
Spin-nematic squeezing occurs at the spectral edges among the retained states for $N=240$ and $q=0.4,0.6,0.8,1.0,1.2$: every state with $\xi_x^2<1$ has $m=0$ or $m=120$. In the middle interval $m=5,\ldots,115$, the smallest retained $\xi_x^2$ is $10.18$, $10.47$, $8.570$, $6.977$, and $5.537$, respectively. Figure~\ref{fig:profiles}(b) contrasts this edge concentration with the QFI dip.

\section{Semiclassical description of the dip}\label{sec:semiclassical}
The normalized dip decreases across the seven sizes $N=60,120,240,480,960,1920,3840$ at each field $q=0.6,0.8,1.0,1.2$; at $q=1$ it falls from $0.6480$ to $0.3686$. To describe this decrease, we write $\epsilon=-q+\delta$ and define
\begin{equation}
 u(\rho)=\frac{-q\rho-\epsilon}{2\rho(1-\rho)},\qquad
 w(\rho)=\frac{1}{\rho(1-\rho)\sqrt{u(1-u)}} .
\end{equation}
Following the semiclassical relation between eigenstate expectations and normalized orbit averages \cite{Engelhardt2015,Ribeiro2008}, we evaluate the average on the intervals with $0<u<1$ as
\begin{equation}
 F_{\rm cl}(q,\delta)=
 2\frac{\int 4\rho(1-\rho)u(\rho)w(\rho)\,d\rho}
          {\int w(\rho)\,d\rho}.
 \label{eq:average}
\end{equation}
The denominator $\int w\,d\rho$ is proportional to the classical density of states and to the orbit period; its logarithmic divergence at $\epsilon_c$ is the ESQPT density-of-states singularity \cite{Feldmann2021}. The numerator $\int4\rho(1-\rho)u w\,d\rho$ remains finite at the polar endpoint $\rho\to1$, suppressing the normalized spin fluctuation $\av{S^2}/N^2$ near the separatrix.
Here $4\rho(1-\rho)\cos^2\phi$ is the classical counterpart of $\av{S^2}/N^2$. Appendix~\ref{app:integrals} gives the endpoint substitution and integration accuracy.

The one-degree-of-freedom logarithmic singularity belongs to the general classification of nondegenerate stationary points \cite{Stransky2016}. Phase-space calculations also reproduce the smoothed level density of Dicke and Tavis--Cummings models \cite{Bastarrachea2014}. In Dicke models, the density of states normalizes microcanonical observable averages \cite{Brandes2013}; such energy-shell averages need not describe each individual eigenstate of a system with several degrees of freedom.

The local dip exponent decreases from $0.137$--$0.238$ across the four fields in the initial size doubling to $0.092$--$0.100$ in the last. It is defined by
\begin{equation}
 e(N)=-\log_2[\fd(2N)/\fd(N)] .
\end{equation}
The full seven-size dip values appear in Appendix~\ref{app:numerics}.

The semiclassical fit describes the six sizes $N\geq120$ with maximum relative residuals $2.8\%$ and $1.5\%$ for positive and negative cutoffs, respectively (Table~\ref{tab:fits}). At each field we fit one parameter $\kappa$ separately, minimizing the maximum relative residual over those six sizes:
\begin{equation}
 r_{\rm sc}^{(\sigma)}=\min_{0.05\leq\kappa\leq2}
 \max_N\left|\frac{F_{\rm cl}(q,\sigma\kappa/N)}{\fd(N,q)}-1\right|,
 \quad \sigma=\pm1 .
\end{equation}
The ratios of the largest to smallest fitted coefficients are $1.140$ and $1.080$, respectively. The positive-cutoff coefficients decrease and then increase across the four fields, while the negative-cutoff coefficients increase.

\par\medskip\noindent\begin{minipage}{\columnwidth}

\captionof{table}{One-parameter fits at the six sizes $N\geq120$: positive and negative cutoffs. Each residual is the maximum relative residual over the six sizes.}\label{tab:fits}
\begin{ruledtabular}\begin{tabular}{rrrrr}
$q$&$\kappa_+$&$r_{\rm sc}^{(+)}$&$\kappa_-$&$r_{\rm sc}^{(-)}$\\
0.6&0.347381&0.024534&0.296742&0.013091\\
0.8&0.329046&0.014660&0.309679&0.012481\\
1.0&0.338572&0.013287&0.316751&0.011402\\
1.2&0.375042&0.027945&0.320400&0.014790
\end{tabular}\end{ruledtabular}

\end{minipage}
\par\medskip

A small power law describes the same six-size data comparably to the positive-cutoff semiclassical fit: $r_P/r_{\rm sc}^{(+)}=0.8803,1.032,1.240,0.7608$ across the four fields. We fit $\ln\fd=a-\gamma\ln N$ by ordinary least squares, using two parameters per field. Table~\ref{tab:allfits} gives $(a,\gamma)$ for $\fd=e^aN^{-\gamma}$ and the single-field coefficients $\kappa_\sigma$ for $F_{\rm cl}(q,\sigma\kappa_\sigma/N)$, $\sigma=\pm1$. Thus both forms give comparable descriptions over $N=120$--$3840$ under their respective fitting losses (Fig.~\ref{fig:scaling}).

Substituting the measured offsets $\delta_{\rm meas}=\epsilon_{\rm dip}+q$ into Eq.~(\ref{eq:average}) gives fieldwise maximum relative residuals of $17.1\%$, $18.8\%$, $17.8\%$, and $29.9\%$ at $q=0.6,0.8,1.0,1.2$. All $24$ measured offsets are negative, and all predictions exceed $\fd$; the resulting $17$--$30\%$ residuals identify $\kappa$ as an effective fitting parameter rather than $N|\delta_{\rm meas}|$. Using the offset $\delta+1/(2N)$ of the Feldmann variable changes the fieldwise maximum absolute relative residuals to $10.85$--$25.73\%$: $25.73\%$, $15.00\%$, $10.85\%$, and $18.77\%$ at the same four fields, respectively. The maximum residual increases at $q=0.6$ and decreases at the other three fields.

To gauge the resolution set by the discrete spectrum, we define $\nu=(F_2-\fd)/\fd$, where $F_2$ is the second-lowest $\F/N^2$ in the window, ordered by fluctuation magnitude. At $21$ of the $24$ fitted points $\nu$ exceeds $|F_{\rm sc}-F_P|/\fd$, the separation between the positive-cutoff semiclassical and power-law predictions; the exceptions are $(q,N)=(0.6,3840),(0.8,480),(1,480)$.

The semiclassical formula extrapolates to a slow $1/\ln N$ decrease when $\delta=\kappa/N$ approaches the separatrix. Its classical values at $q=1$ illustrate this trend:
\begin{equation}
\begin{array}{c|c}
\delta&F_{\rm cl}\\ \hline
10^{-4}&0.3771\\10^{-6}&0.2631\\10^{-8}&0.2019\\
10^{-10}&0.1639\\10^{-12}&0.1378
\end{array}
\end{equation}

\onecolumngrid
\noindent\begin{minipage}{\textwidth}

{\centering\includegraphics[width=0.84\textwidth]{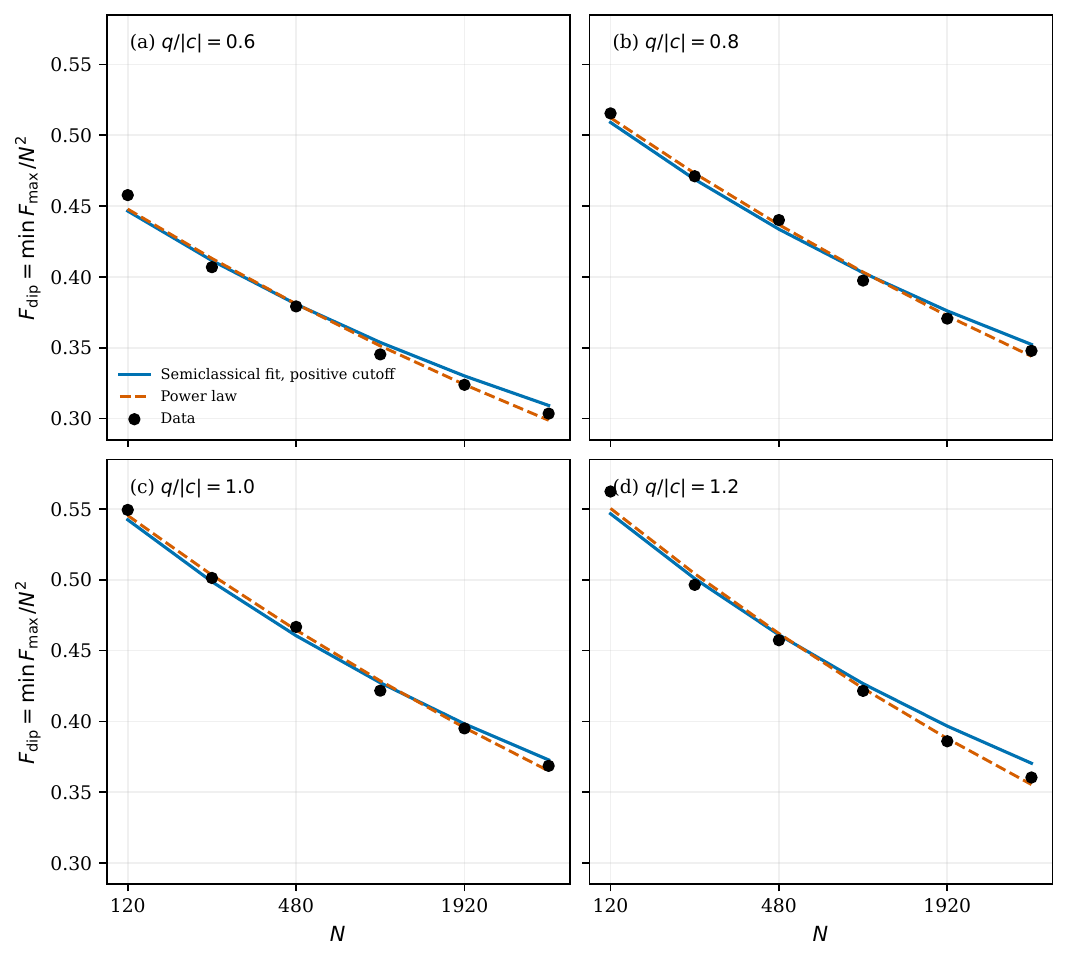}\par}
\captionof{figure}{Normalized dip $\fd=\min_{|\epsilon+q|<0.06}\F/N^2$ for $q=0.6,0.8,1,1.2$ and $N=120,240,480,960,1920,3840$. Solid and dashed curves connect the positive-cutoff semiclassical fit and power-law predictions at the six sizes; the negative-cutoff fits are listed in Table~\ref{tab:fits}.}
\label{fig:scaling}

\end{minipage}
\par\medskip
\twocolumngrid

\section{Slow sweeps through the ESQPT}\label{sec:sweeps}
The excited-state half-excitation times exceed their ground-state counterparts by a factor between $11.2$ and $13.1$ across $N=60,120,240,480$ (Table~\ref{tab:sweep}). We evolve an initial eigenstate $|E_m(q_i)\rangle$ under $q(t)=q_i+(q_f-q_i)t/\tau$ and measure the final excitation probability
\begin{equation}
 \Pe=1-|\av{E_m(q_f)|\psi(\tau)}|^2 .
\end{equation}
The nonzero neighboring couplings of the finite tridiagonal chain give a nondegenerate ordered spectrum, so the same index $m$ specifies the adiabatic target throughout the path. For each finite system, smooth evolution with a nonzero spectral gap permits adiabatic following of this instantaneous eigenspace \cite{Avron1987,Jansen2007}. Ground-state sweeps use $m=0$ and $q:3\to1$. For excited-state sweeps, $m=\mathrm{round}[(N/2+1)/2]$, with halves rounded upward. We locate $q_c$ at the smallest grid value entering the level-count plateau
$\#\{E_j(q)<-qN\}=m$ on a $q$ grid of spacing $0.002$, then choose $q_i=q_c+0.6$ and $q_f=q_c-0.6$. Table~\ref{tab:sweep} gives the paths. The fixed-index targets at negative final fields and the static dip states in positive-field energy windows define two separate problems.

At $\tau=1000$, the excited-state excitation probabilities are $0.00275$, $0.00934$, $0.0194$, and $0.0779$ for $N=60,120,240,480$, while the ground-state values range from $6.33\times10^{-6}$ to $4.76\times10^{-5}$. The durations are $\tau=2,5,10,20,50,100,200,500,1000$. We obtain the half-excitation time $\tau^*$ by linear interpolation of $\Pe$ in $\ln\tau$ at $\Pe=0.5$; both protocols bracket this value at every size.

\onecolumngrid
\noindent\begin{minipage}{\textwidth}

\captionof{table}{Sweep paths and half-excitation times. Ground-state paths are all $3\to1$. The supplemental $N=480$ row is excluded from the exponent fits.}\label{tab:sweep}
\begin{ruledtabular}\begin{tabular}{rrrrrrrrr}
$N$&$m$&$q_c$&$q_i$&$q_f$&$\tau_E^*$&$\tau_g^*$&$\tau_E^*/\tau_g^*$&$\Pe^E(1000)$\\
60&16&0.306&0.906&-0.294&78.382980&6.906613&11.34898&0.002751\\
120&31&0.318&0.918&-0.282&108.289679&8.239752&13.14235&0.009338\\
240&61&0.324&0.924&-0.276&189.522854&14.582867&12.99627&0.019385\\
480&121&0.326&0.926&-0.274&372.862950&33.342873&11.18269&0.077939
\end{tabular}\end{ruledtabular}

\end{minipage}
\par\medskip
\twocolumngrid

The excited-state excitation probability rises over duration intervals $\tau=20\to50$ for $N=60,120$; $2\to5$, $10\to20$, and $50\to100$ for $N=240$; and $10\to20$, $20\to50$, and $50\to100$ for $N=480$ (Fig.~\ref{fig:sweeps}). For example, at $N=240$ it rises from $0.887$ to $0.990$ as $\tau$ increases from $50$ to $100$.

The local size exponents of $\tau^*$ increase from $0.466$ to $0.808$ to $0.976$ for excited-state sweeps and from $0.255$ to $0.824$ to $1.193$ for ground-state sweeps. Log-linear fits over $N=60,120,240$ give $\beta_E=0.637\pm0.098$ and $\beta_g=0.539\pm0.164$, with regression slope standard errors and difference $0.098$. These are finite-size exponents for paths with different initial levels, lengths, and endpoints; the supplemental $N=480$ point displays their continued drift (Fig.~\ref{fig:times}).

\onecolumngrid
\noindent\begin{minipage}{\textwidth}

\includegraphics[width=\textwidth]{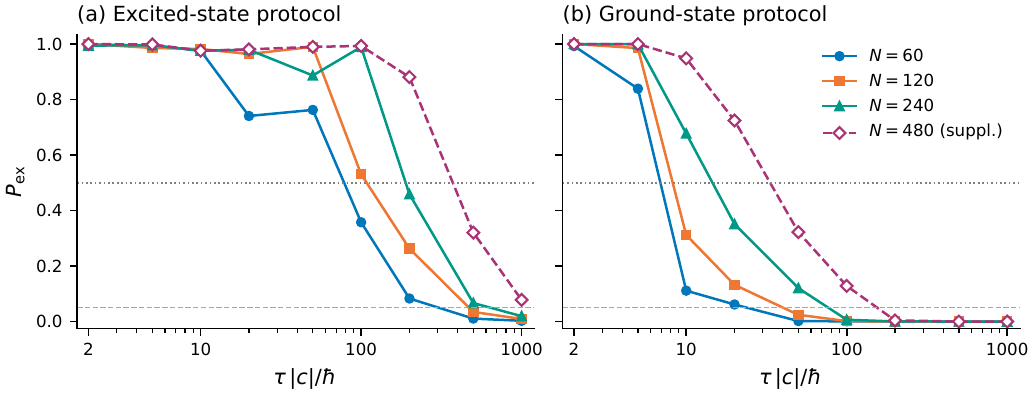}
\captionof{figure}{Excitation probability versus duration for (a) the excited-state paths $q_c(N)+0.6\to q_c(N)-0.6$ and (b) the ground-state path $3\to1$. Symbols are the nine sweep durations for $N=60,120,240,480$. Open diamonds and dashed curves identify the supplemental size. Horizontal lines mark $0.5$ and $0.05$.}
\label{fig:sweeps}

\end{minipage}
\par\medskip
\twocolumngrid

\onecolumngrid
\newpage
\twocolumngrid
\par\medskip\noindent\begin{minipage}{\columnwidth}

\includegraphics[width=\columnwidth]{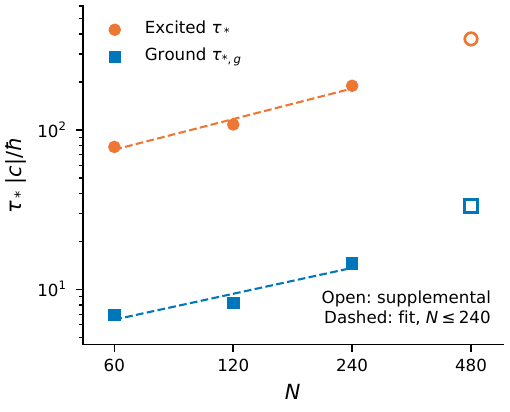}
\captionof{figure}{Half-excitation times. Filled symbols are the primary sizes $N=60,120,240$; open symbols are supplemental $N=480$. Dashed lines span the three-size fitting range. Supplemental points are excluded from both fits.}
\label{fig:times}

\end{minipage}
\par\medskip

\section{Discussion}
We find that the ESQPT appears as a dip in eigenstate rotation fluctuations near the critical energy. At $q=0.6,0.8,1.0,1.2$, the dip value of $\F$ equals $2\av{S^2}$ for the sampled sizes up to $N=3840$, so the feature is a dip in the total-spin fluctuations. Within the semiclassical orbit average with one fitted cutoff $\kappa$ per field, the slow decrease of the normalized dip with $N$ follows from the logarithmic divergence of the classical density of states at the separatrix. Spin-nematic squeezing is concentrated at the spectral edges among the retained states. Separately, fixed-index sweeps that follow an excited level through the ESQPT have half-excitation times $11.2$--$13.1$ times those of the ground-state sweeps.

The dip and the LMG magnetometry peak probe different responses of an eigenstate. Wang and Marzolino \cite{Wang2024} report a peak with fitted scaling near $N^{2.07}$ for the field-dependent Hamiltonian $H_h=hS_z-S_x^2/N$, whose QFI is
\begin{equation}
 {\cal F}_h(E_k)=4\sum_{n\ne k}
 \frac{|\av{E_n|S_z|E_k}|^2}{(E_n-E_k)^2}.
\end{equation}
This observable measures the eigenstate derivative with respect to $h$ through spectral-gap denominators. Our rotation QFI instead measures the maximal variance of a generator in a fixed state [Eq.~(\ref{eq:qfi})]. The two observables therefore give distinct spectral signatures of the ESQPT.

Other eigenstate signatures depend on the observable and spectral averaging. Atomic-inversion profiles resolve excited-state criticality in collective atom--field models; the Dicke comparison averages neighboring eigenstates \cite{PerezFernandez2011}. Smoothed atomic and photon observables also distinguish excited-state phases in the extended Dicke model \cite{Kloc2017}. In two-mode interferometry, metrologically useful entanglement need not imply spin squeezing \cite{Pezze2009}. These precedents motivate specifying the generator and averaging convention when comparing fluctuation signals.

Dicke-model studies of sudden quenches find protocol-dependent stabilization or suppression of the long-time power-law decay stage near an ESQPT \cite{Kloc2018}. These survival dynamics address a different preparation protocol from the fixed-index slow sweeps studied here.

We also examined whether the additional inflection points mark the boundaries of the $\F=2\av{S^2}$ interval. They do so partially: $3$ of $9$ points at $N=240$ and $q=0.6,1.0,1.4$ lie within $0.02$ of a same-field boundary. The interval is the maximal contiguous sequence containing the state nearest $\epsilon=-q$ with $|\F/(2\av{S^2})-1|\leq10^{-6}$; its endpoints are eigenenergies. At $q=1.0$ and $1.4$, the states with the largest absolute second finite difference outside the critical window and spectral endpoints, $m=47$ and $m=50$, are exactly the last interval states. At $q=0.6$, the interval ends at $m=45$ and the strongest additional inflection is at $m=55$, ten levels higher. Across $N=120,240,480$, five of the six boundaries vary by less than $0.02$, while the $q=1.4$ upper boundary shifts by $0.0274$.

\appendix
\section{Matrix elements and operator blocks}\label{app:operators}
The matrix of $S^2$ has diagonal
$2(N-2k)(2k+1)+2k$ and neighboring element
$2(k+1)\sqrt{(N-2k)(N-2k-1)}$.
Writing $a_\pm=a_{\pm1}$, the operators of Ref.~\cite{Huang2022} are
\begin{align}
 S_x&=\frac{a_+^\dagger a_0+a_0^\dagger a_+
                  +a_0^\dagger a_-+a_-^\dagger a_0}{\sqrt2},\\
 Q_{yz}&=\frac{i}{\sqrt2}(a_0^\dagger a_--a_+^\dagger a_0
                         +a_0^\dagger a_+-a_-^\dagger a_0),\\
 Q_{yy}&=\tfrac23n_0-\tfrac13(n_++n_-)
                  -a_+^\dagger a_--a_-^\dagger a_+,\\
 Q_{zz}&=\tfrac23(n_++n_-)-\tfrac43n_0 .
\end{align}
The products $S_x^2,Q_{yz}^2,\{S_x,Q_{yz}\}$ must be formed before projection because the individual operators change $M$. They were constructed in the fixed-$N$ three-mode space of dimension $(N+1)(N+2)/2$ and restricted to the $N/2+1$ dimensional block. The check $S_x^2=S^2/2$ in this block had maximum element discrepancy $4.27\times10^{-14}$ for $N=6,10,20$; the four required blocks were Hermitian. Polar-state normalization and three ground-state comparisons at $N=10$ agreed with independent values within the $10^{-3}$ tolerance (polar $\xi_x^2$ within $10^{-8}$).

The same full-space products were independently evaluated at $N=960,1920,3840$, with dimensions $462241$, $1846081$, and $7378561$. Closed-form basis indexing and sequential release of products reduce storage without projecting away intermediate magnetization sectors. All $2768$ eigenstates in the windows $|\epsilon+q|<0.06$ were measured. At the twelve direct QFI minima, the maximum relative discrepancy from $2\av{S^2}$ is $1.56\times10^{-15}$, below $10^{-8}$, and the proxy and direct minimum select the same index in every case. The direct normalized dips agree with values from the independent $S^2$ calculation within $7.12\times10^{-13}$.

\section{Semiclassical integration and convergence}\label{app:integrals}
For $x=1-\rho$, let $L=2-q$ and define
\begin{align}
 x_0&=\delta/q,\quad
 x_-=\frac{-2\delta}{L+\sqrt{L^2+8\delta}},\\
 x_+&=\frac{L+\sqrt{L^2+8\delta}}4,\\
 l&=\max(x_0,x_-),\qquad o=\min(x_0,x_-).
\end{align}
With $d=l-o$ and $b=x_+-l$, set $x=l+b\sin^2\vartheta$, $0\leq\vartheta\leq\pi/2$. Common factors cancel in Eq.~(\ref{eq:average}), leaving
\begin{align}
 I_0&=\int_0^{\pi/2}\frac{d\vartheta}{\sqrt{d+b\sin^2\vartheta}},\\
 I_1&=\int_0^{\pi/2}\frac{q(l+b\sin^2\vartheta)-\delta}
                       {\sqrt{d+b\sin^2\vartheta}}\,d\vartheta,\\
 F_{\rm cl}&=4I_1/I_0 .
\end{align}
The adaptive quadrature began at relative tolerance $10^{-5}$ and tightened it until the relative changes in numerator, denominator, and ratio were all below $10^{-6}$; the largest accepted change in the fit calculations was $6.73\times10^{-7}$. Transformed-integral values agreed with the reference implementation within $4.70\times10^{-13}$. The five direct-$\rho$ comparisons differed by at most $3.00\times10^{-8}$, within the $10^{-6}$ comparison tolerance. Two independent orbit averages at $q=1$, $\delta=10^{-2},10^{-3}$ agreed within $1.78\times10^{-13}$.

The same endpoint substitution applies to negative $\delta$ by the definitions of $l$ and $o$. The $817$ unique integration arguments from the negative-cutoff fits, the $24$ measured-offset comparisons, and one reference comparison had maximum accepted relative change $1.58\times10^{-8}$. At $q=1$, $\delta=-10^{-3}$, it gives $F_{\rm cl}=0.485486201166$, differing from the reference value by $2.35\times10^{-14}$. The quadratures apply to both cutoff signs.

\section{Numerical details}\label{app:numerics}
The Hamiltonian was checked against full three-mode construction at $N=6,8$ for both signs of $c$, with maximum eigenvalue discrepancy $1.78\times10^{-15}$. Time evolution used DOP853 \cite{Hairer1993} with relative tolerance $10^{-9}$ and absolute tolerance $10^{-11}$, removing a scalar phase and normalizing the final projection. Fourteen ground-state benchmarks at $N=100,200$ had maximum absolute probability deviation $1.10\times10^{-4}$. Halving both tolerances once at $N=200,\tau=20$ changed the probability by $1.67\times10^{-11}$.

\onecolumngrid
\noindent
\begin{minipage}[t]{0.48\textwidth}\small
\captionof{table}{The $28$ normalized dip values $\fd(N,q)$. Printed values are rounded to nine decimals.}\label{tab:dipvalues}
\begin{ruledtabular}\begin{tabular}{rllll}
$N$&$q=0.6$&$q=0.8$&$q=1.0$&$q=1.2$\\
60&0.503422753&0.586902161&0.647950942&0.631307476\\
120&0.457770768&0.515497606&0.549360482&0.562371811\\
240&0.406856184&0.471090584&0.501296103&0.496432658\\
480&0.379144772&0.440138523&0.466717608&0.457318136\\
960&0.345235327&0.397384536&0.421678328&0.421533582\\
1920&0.323762078&0.370545075&0.395029668&0.385940265\\
3840&0.303425858&0.347771551&0.368568182&0.360344465\\
\end{tabular}\end{ruledtabular}
\end{minipage}\hfill
\begin{minipage}[t]{0.48\textwidth}\small
\captionof{table}{Complete parameters for reconstructing the six-size fits. The semiclassical fit is $F_{\rm cl}(q,\sigma\kappa_\sigma/N)$, $\sigma=\pm1$, with $F_{\rm cl}$ defined in Eq.~(\ref{eq:average}); the power law is $\fd=e^aN^{-\gamma}$. The quantities $r_P$ and $r_{\rm sc}^{(\pm)}$ are maximum relative residuals. Printed parameters are rounded to nine decimals.}\label{tab:allfits}
\small\setlength{\tabcolsep}{2pt}
\begin{ruledtabular}\begin{tabular}{rrrrr}
$q$&$\kappa_+$&$r_{\rm sc}^{(+)}$&$\kappa_-$&$r_{\rm sc}^{(-)}$\\
0.6&0.347380555&0.024534279&0.296742168&0.013091187\\
0.8&0.329045762&0.014660435&0.309678712&0.012480696\\
1.0&0.338571678&0.013286600&0.316750541&0.011401997\\
1.2&0.375041552&0.027944604&0.320399755&0.014790067\\
\end{tabular}\end{ruledtabular}
\begin{ruledtabular}\begin{tabular}{rrrr}
$q$&$a$&$\gamma$&$r_P$\\
0.6&-0.243723379&0.116866571&0.021597750\\
0.8&-0.118015935&0.115017637&0.015129801\\
1.0&-0.051272114&0.115903459&0.016473290\\
1.2&0.007229887&0.126227141&0.021260194\\
\end{tabular}\end{ruledtabular}
\end{minipage}
\par\medskip
\twocolumngrid
Table~\ref{tab:dipvalues} lists all $28$ dip values; the fits use their unrounded values. Table~\ref{tab:allfits} gives the positive- and negative-cutoff semiclassical fits together with the power-law fits.

\bibliographystyle{apsrev4-2}
\bibliography{refs}

@article{Feldmann2021,
 author={Feldmann, Polina and Klempt, Carsten and Smerzi, Augusto and Santos, Luis and Gessner, Manuel},
 title={Interferometric Order Parameter for Excited-State Quantum Phase Transitions in {Bose-Einstein} Condensates},
 journal={Phys. Rev. Lett.}, volume={126}, pages={230602}, year={2021},
 doi={10.1103/PhysRevLett.126.230602}, eprint={2011.02823}, archivePrefix={arXiv}}

@article{Caprio2008,
 author={Caprio, M. A. and Cejnar, P. and Iachello, F.},
 title={Excited state quantum phase transitions in many-body systems},
 journal={Ann. Phys.}, volume={323}, number={5}, pages={1106--1135}, year={2008},
 doi={10.1016/j.aop.2007.06.011}, eprint={0707.0325}, archivePrefix={arXiv}}

@article{Cejnar2021,
 author={Cejnar, Pavel and Str{\'a}nsk{\'y}, Pavel and Macek, Michal and Kloc, Michal},
 title={Excited-state quantum phase transitions},
 journal={J. Phys. A: Math. Theor.}, volume={54}, pages={133001}, year={2021},
 doi={10.1088/1751-8121/abdfe8}, eprint={2011.01662}, archivePrefix={arXiv}}

@article{Engelhardt2015,
 author={Engelhardt, G. and Bastidas, V. M. and Kopylov, W. and Brandes, T.},
 title={Excited-state quantum phase transitions and periodic dynamics},
 journal={Phys. Rev. A}, volume={91}, pages={013631}, year={2015},
 doi={10.1103/PhysRevA.91.013631}, eprint={1405.3514}, archivePrefix={arXiv}}

@article{Wang2024,
 author={Wang, Qian and Marzolino, Ugo},
 title={Precision magnetometry exploiting excited state quantum phase transitions},
 journal={SciPost Phys.}, volume={17}, pages={043}, year={2024},
 doi={10.21468/SciPostPhys.17.2.043}, eprint={2306.01126}, archivePrefix={arXiv}}

@article{Huang2022,
 author={Huang, Yixiao and Ding, Yi and Xu, Jiahao and Liu, Jianning and Wang, Hailong and Xiong, Heng-Na},
 title={Spin-nematic squeezing for dynamical quantum phase transitions in a spinor {Bose-Einstein} condensate},
 journal={Phys. Rev. A}, volume={106}, pages={022430}, year={2022},
 doi={10.1103/PhysRevA.106.022430}}

@article{Zhou2023,
 author={Zhou, Lu and Kong, Jia and Lan, Zhihao and Zhang, Weiping},
 title={Dynamical quantum phase transitions in a spinor {Bose-Einstein} condensate and criticality enhanced quantum sensing},
 journal={Phys. Rev. Research}, volume={5}, pages={013087}, year={2023},
 doi={10.1103/PhysRevResearch.5.013087}, eprint={2209.11415}, archivePrefix={arXiv}}

@article{Niu2023,
 author={Niu, Zhen-Xia and Wang, Qian},
 title={Characterizing dynamical phase transitions in a spinor {Bose-Einstein} condensate via quantum and semiclassical analyses},
 journal={Phys. Scr.}, volume={99}, pages={065415}, year={2024},
 doi={10.1088/1402-4896/ad4a9d}, eprint={2310.15841}, archivePrefix={arXiv}}

@article{Xin2023,
 author={Xin, Lin and Barrios, Maryrose and Cohen, Julia T. and Chapman, Michael S.},
 title={Long-Lived Squeezed Ground States in a Quantum Spin Ensemble},
 journal={Phys. Rev. Lett.}, volume={131}, pages={133402}, year={2023},
 doi={10.1103/PhysRevLett.131.133402}, eprint={2202.12338}, archivePrefix={arXiv}}

@article{Xin2022,
 author={Xin, Lin and Chapman, M. S. and Kennedy, T. A. B.},
 title={Fast Generation of Time-Stationary Spin-1 Squeezed States by Nonadiabatic Control},
 journal={PRX Quantum}, volume={3}, pages={010328}, year={2022},
 doi={10.1103/PRXQuantum.3.010328}, eprint={2109.12168}, archivePrefix={arXiv}}

@article{Anquez2016,
 author={Anquez, M. and Robbins, B. A. and Bharath, H. M. and Boguslawski, M. and Hoang, T. M. and Chapman, M. S.},
 title={Quantum {Kibble-Zurek} Mechanism in a Spin-1 {Bose-Einstein} Condensate},
 journal={Phys. Rev. Lett.}, volume={116}, pages={155301}, year={2016},
 doi={10.1103/PhysRevLett.116.155301}, eprint={1512.06914}, archivePrefix={arXiv}}

@article{Luo2017,
 author={Luo, Xin-Yu and Zou, Yi-Quan and Wu, Ling-Na and Liu, Qi and Han, Ming-Fei and Tey, Meng Khoon and You, Li},
 title={Deterministic entanglement generation from driving through quantum phase transitions},
 journal={Science}, volume={355}, pages={620--623}, year={2017},
 doi={10.1126/science.aag1106}, eprint={1702.03120}, archivePrefix={arXiv}}

@article{Braunstein1994,
 author={Braunstein, Samuel L. and Caves, Carlton M.},
 title={Statistical distance and the geometry of quantum states},
 journal={Phys. Rev. Lett.}, volume={72}, pages={3439--3443}, year={1994},
 doi={10.1103/PhysRevLett.72.3439}}

@article{Hamley2012,
 author={Hamley, C. D. and Gerving, C. S. and Hoang, T. M. and Bookjans, E. M. and Chapman, M. S.},
 title={Spin-nematic squeezed vacuum in a quantum gas},
 journal={Nat. Phys.}, volume={8}, pages={305--308}, year={2012},
 doi={10.1038/nphys2245}, eprint={1111.1694}, archivePrefix={arXiv}}

@article{Huang2015,
 author={Huang, Yixiao and Xiong, Heng-Na and Sun, Zhe and Wang, Xiaoguang},
 title={Generation and storage of spin-nematic squeezing in a spinor {Bose-Einstein} condensate},
 journal={Phys. Rev. A}, volume={92}, pages={023622}, year={2015},
 doi={10.1103/PhysRevA.92.023622}}

@article{Hoang2016,
 author={Hoang, Thai M. and Bharath, Hebbe M. and Boguslawski, Matthew J. and Anquez, Martin and Robbins, Bryce A. and Chapman, Michael S.},
 title={Adiabatic quenches and characterization of amplitude excitations in a continuous quantum phase transition},
 journal={Proc. Natl. Acad. Sci. U.S.A.}, volume={113}, number={34}, pages={9475--9479}, year={2016},
 doi={10.1073/pnas.1600267113}, eprint={1512.06766}, archivePrefix={arXiv}}

@article{Niu2024,
 author={Niu, Zhen-Xia and Wang, Qian},
 title={Multifractality and excited-state quantum phase transition in ferromagnetic spin-1 {Bose-Einstein} condensates},
 journal={Phys. Rev. E}, volume={110}, pages={064112}, year={2024},
 doi={10.1103/PhysRevE.110.064112}, eprint={2407.20612}, archivePrefix={arXiv}}

@article{MeyerHoppe2023,
 author={Meyer-Hoppe, B. and Anders, F. and Feldmann, P. and Santos, L. and Klempt, C.},
 title={Excited-State Phase Diagram of a Ferromagnetic Quantum Gas},
 journal={Phys. Rev. Lett.}, volume={131}, pages={243402}, year={2023},
 doi={10.1103/PhysRevLett.131.243402}, eprint={2301.10655}, archivePrefix={arXiv}}

@article{Jansen2007,
 author={Jansen, Sabine and Ruskai, Mary-Beth and Seiler, Ruedi},
 title={Bounds for the adiabatic approximation with applications to quantum computation},
 journal={J. Math. Phys.}, volume={48}, pages={102111}, year={2007},
 doi={10.1063/1.2798382}, eprint={quant-ph/0603175}, archivePrefix={arXiv}}

@book{Hairer1993,
 author={Hairer, Ernst and N{\o}rsett, Syvert P. and Wanner, Gerhard},
 title={Solving Ordinary Differential Equations {I}: Nonstiff Problems},
 edition={2}, publisher={Springer}, address={Berlin}, year={1993},
 series={Springer Series in Computational Mathematics}, volume={8},
 doi={10.1007/978-3-540-78862-1}}

@article{Ho1998,
 author={Ho, Tin-Lun},
 title={Spinor {Bose} Condensates in Optical Traps},
 journal={Phys. Rev. Lett.}, volume={81}, pages={742--745}, year={1998},
 doi={10.1103/PhysRevLett.81.742}, eprint={cond-mat/9803231}, archivePrefix={arXiv}}

@article{Ohmi1998,
 author={Ohmi, Tetsuo and Machida, Kazushige},
 title={{Bose--Einstein} Condensation with Internal Degrees of Freedom in Alkali Atom Gases},
 journal={J. Phys. Soc. Jpn.}, volume={67}, pages={1822--1825}, year={1998},
 doi={10.1143/JPSJ.67.1822}, eprint={cond-mat/9803160}, archivePrefix={arXiv}}

@article{Law1998,
 author={Law, C. K. and Pu, H. and Bigelow, N. P.},
 title={Quantum Spins Mixing in Spinor {Bose--Einstein} Condensates},
 journal={Phys. Rev. Lett.}, volume={81}, pages={5257--5261}, year={1998},
 doi={10.1103/PhysRevLett.81.5257}, eprint={cond-mat/9807258}, archivePrefix={arXiv}}

@article{Yi2002,
 author={Yi, S. and M{\"u}stecapl{\i}o{\u g}lu, {\"O}. E. and Sun, C. P. and You, L.},
 title={Single-mode approximation in a spinor-1 atomic condensate},
 journal={Phys. Rev. A}, volume={66}, pages={011601}, year={2002},
 doi={10.1103/PhysRevA.66.011601}, eprint={cond-mat/0201173}, archivePrefix={arXiv}}

@article{Zhang2005,
 author={Zhang, Wenxian and Zhou, D. L. and Chang, M.-S. and Chapman, M. S. and You, L.},
 title={Coherent spin mixing dynamics in a spin-1 atomic condensate},
 journal={Phys. Rev. A}, volume={72}, pages={013602}, year={2005},
 doi={10.1103/PhysRevA.72.013602}, eprint={cond-mat/0502061}, archivePrefix={arXiv}}

@article{Kawaguchi2012,
 author={Kawaguchi, Yuki and Ueda, Masahito},
 title={Spinor {Bose--Einstein} condensates},
 journal={Phys. Rep.}, volume={520}, pages={253--381}, year={2012},
 doi={10.1016/j.physrep.2012.07.005}, eprint={1001.2072}, archivePrefix={arXiv}}

@article{Chang2005,
 author={Chang, Ming-Shien and Qin, Qishu and Zhang, Wenxian and You, Li and Chapman, Michael S.},
 title={Coherent spinor dynamics in a spin-1 {Bose} condensate},
 journal={Nat. Phys.}, volume={1}, pages={111--116}, year={2005},
 doi={10.1038/nphys153}, eprint={cond-mat/0509341}, archivePrefix={arXiv}}

@article{Sadler2006,
 author={Sadler, L. E. and Higbie, J. M. and Leslie, S. R. and Vengalattore, M. and Stamper-Kurn, D. M.},
 title={Spontaneous symmetry breaking in a quenched ferromagnetic spinor {Bose--Einstein} condensate},
 journal={Nature}, volume={443}, pages={312--315}, year={2006},
 doi={10.1038/nature05094}, eprint={cond-mat/0605351}, archivePrefix={arXiv}}

@article{Black2007,
 author={Black, A. T. and Gomez, E. and Turner, L. D. and Jung, S. and Lett, P. D.},
 title={Spinor Dynamics in an Antiferromagnetic Spin-1 Condensate},
 journal={Phys. Rev. Lett.}, volume={99}, pages={070403}, year={2007},
 doi={10.1103/PhysRevLett.99.070403}, eprint={0704.0925}, archivePrefix={arXiv}}

@article{Gerving2012,
 author={Gerving, C. S. and Hoang, T. M. and Land, B. J. and Anquez, M. and Hamley, C. D. and Chapman, M. S.},
 title={Non-equilibrium dynamics of an unstable quantum pendulum explored in a spin-1 {Bose--Einstein} condensate},
 journal={Nat. Commun.}, volume={3}, pages={1169}, year={2012},
 doi={10.1038/ncomms2179}, eprint={1205.2121}, archivePrefix={arXiv}}

@article{Tian2020,
 author={Tian, T. and Yang, H.-X. and Qiu, L.-Y. and Liang, H.-Y. and Yang, Y.-B. and Xu, Y. and Duan, L.-M.},
 title={Observation of Dynamical Quantum Phase Transitions with Correspondence in an Excited State Phase Diagram},
 journal={Phys. Rev. Lett.}, volume={124}, pages={043001}, year={2020},
 doi={10.1103/PhysRevLett.124.043001}, eprint={2001.02686}, archivePrefix={arXiv}}

@article{Leyvraz2005,
  author = {Leyvraz, F. and Heiss, W. D.},
  title = {Large-{$N$} Scaling Behavior of the {Lipkin-Meshkov-Glick} Model},
  journal = {Phys. Rev. Lett.},
  volume = {95},
  pages = {050402},
  year = {2005},
  doi = {10.1103/PhysRevLett.95.050402},
  eprint = {quant-ph/0507004},
  archivePrefix = {arXiv}
}

@article{Cejnar2008,
  author = {Cejnar, Pavel and Str{\'a}nsk{\'y}, Pavel},
  title = {Impact of quantum phase transitions on excited-level dynamics},
  journal = {Phys. Rev. E},
  volume = {78},
  pages = {031130},
  year = {2008},
  doi = {10.1103/PhysRevE.78.031130},
  eprint = {0807.3649},
  archivePrefix = {arXiv}
}

@article{Stransky2014,
  author = {Str{\'a}nsk{\'y}, Pavel and Macek, Michal and Cejnar, Pavel},
  title = {Excited-state quantum phase transitions in systems with two degrees of freedom: Level density, level dynamics, thermal properties},
  journal = {Ann. Phys.},
  volume = {345},
  pages = {73--97},
  year = {2014},
  doi = {10.1016/j.aop.2014.03.006}
}

@article{Santos2015,
  author = {Santos, Lea F. and P{\'e}rez-Bernal, Francisco},
  title = {Structure of eigenstates and quench dynamics at an excited-state quantum phase transition},
  journal = {Phys. Rev. A},
  volume = {92},
  pages = {050101},
  year = {2015},
  doi = {10.1103/PhysRevA.92.050101},
  eprint = {1506.06765},
  archivePrefix = {arXiv}
}

@article{PerezFernandez2011,
  author = {P{\'e}rez-Fern{\'a}ndez, P. and Rela{\~n}o, A. and Arias, J. M. and Cejnar, P. and Dukelsky, J. and Garc{\'i}a-Ramos, J. E.},
  title = {Excited-state phase transition and onset of chaos in quantum optical models},
  journal = {Phys. Rev. E},
  volume = {83},
  pages = {046208},
  year = {2011},
  doi = {10.1103/PhysRevE.83.046208}
}

@article{Kloc2017,
  author = {Kloc, Michal and Str{\'a}nsk{\'y}, Pavel and Cejnar, Pavel},
  title = {Quantum phases and entanglement properties of an extended {Dicke} model},
  journal = {Ann. Phys.},
  volume = {382},
  pages = {85--111},
  year = {2017},
  doi = {10.1016/j.aop.2017.04.005},
  eprint = {1609.02758},
  archivePrefix = {arXiv}
}

@article{Kloc2018,
  author = {Kloc, Michal and Str{\'a}nsk{\'y}, Pavel and Cejnar, Pavel},
  title = {Quantum quench dynamics in {Dicke} superradiance models},
  journal = {Phys. Rev. A},
  volume = {98},
  pages = {013836},
  year = {2018},
  doi = {10.1103/PhysRevA.98.013836},
  eprint = {1805.06285},
  archivePrefix = {arXiv}
}

@article{Ribeiro2008,
  author = {Ribeiro, Pedro and Vidal, Julien and Mosseri, R{\'e}my},
  title = {Exact spectrum of the {Lipkin-Meshkov-Glick} model in the thermodynamic limit and finite-size corrections},
  journal = {Phys. Rev. E},
  volume = {78},
  pages = {021106},
  year = {2008},
  doi = {10.1103/PhysRevE.78.021106},
  eprint = {0805.4078},
  archivePrefix = {arXiv}
}

@article{Stransky2016,
  author = {Str{\'a}nsk{\'y}, Pavel and Cejnar, Pavel},
  title = {Classification of excited-state quantum phase transitions for arbitrary number of degrees of freedom},
  journal = {Phys. Lett. A},
  volume = {380},
  pages = {2637--2643},
  year = {2016},
  doi = {10.1016/j.physleta.2016.06.031},
  eprint = {1606.06631},
  archivePrefix = {arXiv}
}

@article{Brandes2013,
  author = {Brandes, Tobias},
  title = {Excited-state quantum phase transitions in {Dicke} superradiance models},
  journal = {Phys. Rev. E},
  volume = {88},
  pages = {032133},
  year = {2013},
  doi = {10.1103/PhysRevE.88.032133},
  eprint = {1306.6435},
  archivePrefix = {arXiv}
}

@article{Bastarrachea2014,
  author = {Bastarrachea-Magnani, M. A. and Lerma-Hern{\'a}ndez, S. and Hirsch, J. G.},
  title = {Comparative quantum and semiclassical analysis of atom-field systems. {I}. Density of states and excited-state quantum phase transitions},
  journal = {Phys. Rev. A},
  volume = {89},
  pages = {032101},
  year = {2014},
  doi = {10.1103/PhysRevA.89.032101},
  eprint = {1312.2665},
  archivePrefix = {arXiv}
}

@article{Kitagawa1993,
 author={Kitagawa, Masahiro and Ueda, Masahito},
 title={Squeezed spin states},
 journal={Phys. Rev. A}, volume={47}, number={6}, pages={5138--5143}, year={1993},
 doi={10.1103/PhysRevA.47.5138}}

@article{Wineland1994,
 author={Wineland, D. J. and Bollinger, J. J. and Itano, W. M. and Heinzen, D. J.},
 title={Squeezed atomic states and projection noise in spectroscopy},
 journal={Phys. Rev. A}, volume={50}, number={1}, pages={67--88}, year={1994},
 doi={10.1103/PhysRevA.50.67}}

@article{Pezze2009,
 author={Pezz{\'e}, Luca and Smerzi, Augusto},
 title={Entanglement, Nonlinear Dynamics, and the {Heisenberg} Limit},
 journal={Phys. Rev. Lett.}, volume={102}, number={10}, pages={100401}, year={2009},
 doi={10.1103/PhysRevLett.102.100401}, eprint={0711.4840}, archivePrefix={arXiv}}

@article{Hyllus2012,
 author={Hyllus, Philipp and Laskowski, Wies{\l}aw and Krischek, Roland and Schwemmer, Christian and Wieczorek, Witlef and Weinfurter, Harald and Pezz{\'e}, Luca and Smerzi, Augusto},
 title={Fisher information and multiparticle entanglement},
 journal={Phys. Rev. A}, volume={85}, number={2}, pages={022321}, year={2012},
 doi={10.1103/PhysRevA.85.022321}, eprint={1006.4366}, archivePrefix={arXiv}}

@article{Yukawa2013,
 author={Yukawa, Emi and Ueda, Masahito and Nemoto, Kae},
 title={Classification of spin-nematic squeezing in spin-1 collective atomic systems},
 journal={Phys. Rev. A}, volume={88}, number={3}, pages={033629}, year={2013},
 doi={10.1103/PhysRevA.88.033629}, eprint={1305.2560}, archivePrefix={arXiv}}

@article{Gabbrielli2015,
 author={Gabbrielli, Marco and Pezz{\`e}, Luca and Smerzi, Augusto},
 title={Spin-Mixing Interferometry with {Bose-Einstein} Condensates},
 journal={Phys. Rev. Lett.}, volume={115}, number={16}, pages={163002}, year={2015},
 doi={10.1103/PhysRevLett.115.163002}, eprint={1503.08582}, archivePrefix={arXiv}}

@article{Lucke2011,
 author={L{\"u}cke, B. and Scherer, M. and Kruse, J. and Pezz{\'e}, L. and Deuretzbacher, F. and Hyllus, P. and Topic, O. and Peise, J. and Ertmer, W. and Arlt, J. and Santos, L. and Smerzi, A. and Klempt, C.},
 title={Twin Matter Waves for Interferometry Beyond the Classical Limit},
 journal={Science}, volume={334}, number={6057}, pages={773--776}, year={2011},
 doi={10.1126/science.1208798}, eprint={1204.4102}, archivePrefix={arXiv}}

@article{Zou2018,
 author={Zou, Yi-Quan and Wu, Ling-Na and Liu, Qi and Luo, Xin-Yu and Guo, Shuai-Feng and Cao, Jia-Hao and Tey, Meng Khoon and You, Li},
 title={Beating the classical precision limit with spin-1 {Dicke} states of more than 10,000 atoms},
 journal={Proc. Natl. Acad. Sci. U.S.A.}, volume={115}, number={25}, pages={6381--6385}, year={2018},
 doi={10.1073/pnas.1715105115}, eprint={1802.10288}, archivePrefix={arXiv}}

@article{Avron1987,
 author={Avron, J. E. and Seiler, R. and Yaffe, L. G.},
 title={Adiabatic theorems and applications to the quantum {Hall} effect},
 journal={Commun. Math. Phys.}, volume={110}, pages={33--49}, year={1987},
 doi={10.1007/BF01209015},
 note={Erratum: Commun. Math. Phys. 156, 649--650 (1993)}}

@article{Zurek2005,
 author={Zurek, Wojciech H. and Dorner, Uwe and Zoller, Peter},
 title={Dynamics of a Quantum Phase Transition},
 journal={Phys. Rev. Lett.}, volume={95}, pages={105701}, year={2005},
 doi={10.1103/PhysRevLett.95.105701}, eprint={cond-mat/0503511}, archivePrefix={arXiv}}

@article{Polkovnikov2005,
 author={Polkovnikov, Anatoli},
 title={Universal adiabatic dynamics in the vicinity of a quantum critical point},
 journal={Phys. Rev. B}, volume={72}, pages={161201}, year={2005},
 doi={10.1103/PhysRevB.72.161201}, eprint={cond-mat/0312144}, archivePrefix={arXiv}}

@article{Dziarmaga2005,
 author={Dziarmaga, Jacek},
 title={Dynamics of a Quantum Phase Transition: Exact Solution of the Quantum {Ising} Model},
 journal={Phys. Rev. Lett.}, volume={95}, pages={245701}, year={2005},
 doi={10.1103/PhysRevLett.95.245701}, eprint={cond-mat/0509490}, archivePrefix={arXiv}}

@article{Damski2007,
 author={Damski, Bogdan and Zurek, Wojciech H.},
 title={Dynamics of a Quantum Phase Transition in a Ferromagnetic {Bose-Einstein} Condensate},
 journal={Phys. Rev. Lett.}, volume={99}, pages={130402}, year={2007},
 doi={10.1103/PhysRevLett.99.130402}, eprint={0704.0440}, archivePrefix={arXiv}}

@article{Zhang2013,
 author={Zhang, Z. and Duan, L.-M.},
 title={Generation of Massive Entanglement through an Adiabatic Quantum Phase Transition in a Spinor Condensate},
 journal={Phys. Rev. Lett.}, volume={111}, pages={180401}, year={2013},
 doi={10.1103/PhysRevLett.111.180401}, eprint={1308.2318}, archivePrefix={arXiv}}
\end{document}